\documentclass{aastex631}
\usepackage{amssymb, amsmath}
\usepackage{array} 
\usepackage{geometry}
\usepackage{subfigure}
\usepackage{graphicx}
\usepackage{hyperref}

\begin{document}
\title{Predictions of Transiting Exoplanet Confirmations from Rubin LSST Surveys}
\author[0009-0006-2305-7183]{Corley, S.}
\affiliation{School of Earth and Space Exploration, Arizona State University, 781 E Terrace Mall, Tempe, AZ 85287-6004, United States}
\author[0000-0002-5077-6734]{Feigelson, E.D.}
\affiliation{Departments of Astronomy \& Astrophysics and of Statistics
Penn State University, University Park PA 16801} 
\author[0000-0002-6617-3823]{Caceres, C.}
\affiliation{Instituto de Astrofisica, Departamento de Fisica y Astronomia, Facultad de Ciencias Exactas, Universidad Andres Bello, Fernandez Concha 700, Santiago, Chile}
\author[0000-0002-1046-1500]{Cuevas-Otahola, B.}
\affiliation{Instituto de Astronom\'ia, Universidad Nacional Autonoma de M\'exico, Ciudad Universitaria, Coyoac\'an, CP 04510, Ciudad de M\'exico, M\'exico}
\author[0000-0001-5139-1978]{Kova{\v c}evi{\'c}, A.B.}
\affiliation{Department of Astronomy, Faculty of Mathematics, University of Belgrade, Studentski trg 16, 11000 Belgrade, Serbia}
\date{March 2026}

\begin{abstract} 
\label{abstract}
We assess the prospects for exoplanet transit observations in the 10-year Wide Fast Deep (WFD) and Deep Drilling Field (DDF) surveys within the Legacy Survey of Space and Time (LSST) mission of the Vera C. Rubin Observatory.  We construct a framework for systematic assessment of expected exoplanet yields, highlighting the principal limitations imposed by the survey observing strategy and cadence. We simulate light curves with a wide range of exoplanetary system models derived from planet occurrence rates developed with data from the Kepler mission.  Transit counts for the stellar population are calculated using the TRILEGAL Galactic structure model, incorporating telescope sensitivity and survey cadences.  We apply the constraints that the full duration of at least three transits must be observed and that the signal-to-noise ratio will support detectability. The observations were then validated using the Transit Least Squares periodogram.  

Our findings indicate a limited potential for exoplanet confirmations under the current survey design. Only a small number of hot planets orbiting faint M class main sequence stars will be confirmed in the DDF fields.  The WFD survey is projected to produce no confirmations. These findings underscore the constraints imposed by the sparse, multi-band observing strategy, which prioritizes cosmology and extragalactic science over the continuous photometric coverage required for confirmations.   
\end{abstract}

\keywords{Exoplanets - Irregular cadence - M dwarf stars - Sky surveys - Time series analysis - Transit photometry} % from https://astrothesaurus.org/thesaurus/hierarchical-browse/

\section{Introduction}
\label{intro}
The confirmed exoplanet catalog maintained by the NASA Exoplanet Archive \citep{christiansen2025} now exceeds 6{,}000 entries, yet it does not constitute a census of the Milky Way's exoplanet population. The archive is the product of a small number of targeted missions with narrow fields of view, specific stellar magnitude ranges, and largely uniform cadences. Approximately 90\% of confirmed planets were detected by space-based photometric surveys -- principally \textit{Kepler}, \textit{K2}, and \textit{TESS} \citep{christiansen2025} -- which collectively sampled only a modest fraction of Galactic stellar diversity in terms of spectral type, metallicity, age, and distance. The result is a catalog that is deep within its sampled volumes but highly incomplete across the Galaxy as a whole. Estimates of the total number of exoplanets in the Milky Way consequently span a wide range, from tens of billions to more than a trillion, depending on assumptions about occurrence rates and the stellar populations to which they are extrapolated \citep[e.g.,][]{dressing2015, hsu2019,Bryson2021,Howard2012}.

Extending the observational foundation of this census requires surveys that probe stellar populations and sky regions underrepresented in existing data. Ground-based photometric surveys have contributed meaningfully to the archive: wide-field networks such as SuperWASP \citep{Christian2006} and HATNet \citep{zhang2016} have confirmed hundreds of planets, and the Zwicky Transient Facility (ZTF) has both identified transiting objects around specific stellar classes \citep{parsons2025} and aided confirmation of \textit{TESS} candidates \citep{ross2024}. Nevertheless, ground-based surveys have been fundamentally constrained by atmospheric noise, diurnal gaps, and limited photometric precision, preventing them from meaningfully expanding the parameter space of the confirmed catalog.

A more capable ground-based facility could alter this picture. The Vera C.\ Rubin Observatory's Legacy Survey of Space and Time (Rubin/LSST) will survey $\sim$18{,}000 deg$^2$ of sky over 10 years with a collecting area and photometric depth substantially exceeding those of any preceding ground-based survey \citep{ivesic2019}. In principle, this reach could extend transit searches to stellar populations at kiloparsec distances and across a far broader range of Galactic environments than is accessible to space-based missions with smaller fields of view. Whether the Rubin/LSST observing strategy --- optimized for cosmological transient science rather than stellar photometry --- can support effective transit detection is the central question this paper addresses.

Initial feasibility studies \citep{lund2015, lund2016} estimated that Rubin/LSST could detect $\sim 10^4-10^5$ transit events during its survey. Subsequent analysis \citep{jacklin2015, jacklin2017} suggested that substantial numbers of exoplanets could be detected even within the initial years of the survey, particularly in the DDF.

However, recently settled decisions \citep{PSTN-056, bianco2022} for the Rubin/LSST operational cadence require a reassessment of its exoplanet detection capabilities . Approximately 99\% \citep{PSTN-056} of the observation time will be devoted to the WFD and DDF surveys.   Under the current plan, WFD fields will be observed twice a night with a 33-minute interval, repeating every three days for 10 years \citep{PSTN-056}.  However, the observations will be divided into 6 photometric bands, complicating the accurate measurements needed for transit studies.  DDF fields will receive approximately 20 exposures in one hour with a similar three-day cadence \citep{lsstsciencebook2009,PSTN-056}.

In this study, we model exoplanet detectability in both WFD and DDF scenarios using updated cadence strategies and established planet occurrence rates from the Kepler mission \citep{dressing2015, hsu2019}. Our aim is to estimate the yield and confirmation probability of transiting exoplanets under realistic observing conditions, and to identify the parameter space in which Rubin/LSST can most effectively contribute to exoplanet science.

Our analysis comprises five steps: assessing expected photometric uncertainties in Rubin/LSST data; generating light curves for synthetic stellar systems; estimating detection rates at a chosen signal-to-noise ratio threshold; combining detection probabilities with published exoplanet occurrence rates; and testing the efficacy of standard exoplanet detection techniques.  Our results not only inform expectations for exoplanet discovery with Rubin/LSST, but also provide guidance for possible future microsurveys aimed at improving exoplanet samples. 

\section{Simulations}
\label{methods}
Our methodology addresses two main objectives: assessing Rubin/LSST's capability to collect data suitable for identifying exoplanets orbiting stars in our galaxy; and evaluating techniques for successfully recovering exoplanet transits from simulated data.  Up to 400 thousand simulated light curves are drawn from up to $\sim$2 million models.   

\subsection{LSST Survey Characteristics}
\label{sec:LSSTsurveys}
The Rubin/LSST WFD and DDF surveys will collect data in six photometric filters (\textit{u, g, r, i, z,} and \textit{y}) with single-visit photometry archived in its database. Our analysis focuses on the \textit{g}-band, which offers a balance between stellar brightness and photometric precision for a broad range of potential host stars. Table~\ref{tab:rubinerr} presents the \textit{g}-band photometric uncertainties, derived following the techniques described by \citet{lund2015}, \citet{ivezic2009}, and \citet{ivesic2019}.

We construct a grid of synthetic exoplanetary systems, each defined by combinations of the host star properties of type and distance and the exoplanet properties of radii and orbital period. Table~\ref{tab:model2} summarizes the ranges of parameters used. 

Several simplified assumptions are made: (i) stars lie on the main sequence; (ii) stars do not have binary companions; (iii) stellar luminosities are constant without variations from magnetic activity; (iv) a single planet in a circular orbit is present; and (v) transit ingress and egress transitions are instantaneous.  The effect of these assumptions is to increase the ability of LSST to detect transiting exoplanets.  Our calculations of transit confirmations are therefore optimistic, and in all likelihood the return from the WFD and DDF will be lower than values obtained here. 

\begin{deluxetable}{cccccccccc}[h]
\centering
\tablecaption{Rubin Single-Visit g-Band Uncertainty \label{tab:rubinerr}}
\tablehead{
\colhead{Mag} & \colhead{$\sigma$} &&& \colhead{Mag} & \colhead{$\sigma$} &&& 
\colhead{Mag} & \colhead{$\sigma$}}
\startdata
17 & 0.00088 & & &  20 & 0.00418  & & & 23 & 0.03974 \\ 
18 & 0.00142 & & &  21 & 0.00805  & & & 24 & 0.09615  \\ 
19 & 0.00237 & & &  22 & 0.01719  & & & 25 & 0.23777 \\ 
\enddata
\tablecomments{These values were derived using the techniques  described by \citet{lund2015}, \citet{ivezic2009}, and \citet{ivesic2019}. The uncertainty for Mag 25 is at the survey detection limit for Rubin/LSST. }
\end{deluxetable}

\begin{deluxetable}{lll}
\tablecaption{Model Parameters for Sensitivity Analysis} 
\label{tab:model2}
\tablehead{
\colhead{Category} & \colhead{Parameter} & \colhead{Values}
}
\startdata
\multicolumn{3}{l}{\textbf{Exoplanet Attributes}} \\
& Type & incremental values from 0.75 to 16 Earth radii \\
& Radius ($R_\oplus$) & Ranges scaled from Type and Earth's radius (solar units $9.16 \times 10^{-3}$) \\
\\
\multicolumn{3}{l}{\textbf{Distance Attributes (parsecs)}} \\
& Observing Distance & 10, 25, 40, 100, 1000, 10000 \\
\\
\multicolumn{3}{l}{\textbf{Orbital Periods (days)}} \\
& Period & Based upon recent TESS results (see \cite{christiansen2025}) \\
\\
\multicolumn{3}{l}{\textbf{Mission Attributes}} \\
& Length (days) & 3650 \\
& Nightly on-sky time (hours) & 6.5 \\
& Stellar populations WFD & F/G/K: 3 billion; M: 430 million \\
& Stellar populations DDF & F/G/K: 400 thousand; M: 200 thousand \\
\\
\multicolumn{3}{l}{\textbf{Survey Configurations}} \\
& WFD: & 33 min between snaps, 2 snaps, 3-day cadence \\
& DDF: & 0.5 min between snaps, 20 snaps, 3-day cadence \\
\enddata
\tablecomments{ Ranges and discrete values of parameters used to construct synthetic exoplanetary systems for sensitivity analysis. These parameters, including stellar class, planetary radius, orbital period, and observing distance, define the parameter space explored in our transit detection simulations. Note that parameter values are not continuous but are sampled from the listed discrete sets provided by \cite{dressing2015} and \cite{hsu2019}. }
\end{deluxetable}

\subsection{Host Stellar Populations} \label{sec:star_pop}
Host stars with spectral types F-G-K and M are defined by the main sequence properties taken from the latest version of the values originally derived in \cite{Petigura2013}.\footnote{URL: $https://www.pas.rochester.edu/~emamajek/EEM\_dwarf\_UBVIJHK\_colors\_Teff.txt$.} Population counts by spectral types and distances were derived from the planned WFD and DDF pointings \citep{Weston2025} as applied to the TRILEGAL simulation of Galactic stars \citep{tio2022}. The TRILEGAL simulation was chosen to provide a real number basis for observation estimates for the WFD and DDF pointings. Database queries of the TRILEGAL simulation for the LSST survey regions were made for magnitude depths of 17.0 through 25.0.  These TRILEGAL queries return $\sim$3.7 billion main sequence stars for the WFD survey and $\sim$0.5 million for the DDF survey. 

\subsection{Planet Populations} \label{sec:planet_pop}
Planet occurrences are associated with stellar populations covered by the WFD and DDF surveys described in \S\ref{sec:star_pop}.  For solar-like F-, G- and K-class    stars, \cite{hsu2019} gives the planetary occurrence rates for planet sizes ranging from inflated Jupiters (16~R$_\oplus$) to sub-Earths (0.75~R$_\oplus$) and orbital periods ranging from ultra short period planets (1~day) to Earth-like orbits (500~day).  These are obtained from the analysis of Kepler confirmed planets by \citet{hsu2019} taking into account geometric, instrumental, and transit detection limitations.  \cite{dressing2015} gives similar planet occurrence rates around M dwarf stars. Here, planetary radii  are restricted to $R \leq 4$~R$_\oplus$.   We note that the reliability of these occurrence rates is uncertain as various research groups have reported a range of rates \citep[e.g.][]{Petigura2013}. 

\subsection{Transit Observation Probabilities}
\label{sec:probabilities}
For each synthetic star-planet system, we generate light curves representing the expected photometric time series. Transit events are modeled as periodic decreases in stellar flux with transit depth calculated as $(R_p/R_\star)^2$. Photometric noise is simulated as additive Gaussian noise with the standard deviation set by the expected Rubin/LSST single-visit photometric uncertainty in the \textit{g}-band, as a function of the star's apparent magnitude (Table~\ref{tab:rubinerr}). 

Most planetary orbits do not produce transits due to geometric constraints \citep[e.g.][]{winn2010}.  The probability of transit is 
        \begin{equation}
        p_{\text{transit}} = \frac{R_* + R_p}{a}
        \end{equation}
where $R_{\star}$ is the stellar radius, $R_p$ is the radius of the planet, and $a$ is the semi-major axis.  If a transit occurs across the middle of the stellar disk, the transit duration for a circular orbit is
         \begin{equation}
        t_{transit} = \frac{R_{\star} P_{\text{orb}}}{\pi a}
        \end{equation}
where  $P_{\text{orb}}$ is the orbital period.
With $t_{exp}$ defined as the difference between the start time of the first exposure and the end time of the last exposure of an observed position and 
$\Delta t_{\rm cad}$ defined as the observational cadence, the probability per-night of observing a transit is 

    \begin{equation} \label{eqn:transprob}
    p_{\text{per-night}} = \frac{\max(0, t_{\rm exp} - t_{transit})}{\Delta t_{\rm cad}} \times p_{\text{transit}}
    \end{equation}.
The probability of multiple transits, $k$ transits in $N$ nights, is given by the binomial distribution        
    \begin{equation}  \label{eqn:transprobk}
    p(\text{at least } k) = \sum_{i=k}^N \binom{N}{i} (p_{\text{per-night}})^i (1 - p_{\text{per-night}})^{N - i}
    \end{equation}

\subsection{Combining Stellar and Planetary Populations with Transit Probabilities}
For each stellar class (F, G, K, and M), we constructed a grid that spans the planetary radius and orbital period bins following the organization used by \cite{hsu2019} and \cite{dressing2015}.  We distribute TRILEGAL stellar counts into the occurrence rate grid using the following weighting scheme:
\begin{equation} \label{eqn:popweighted}
f_{i,j} = F_{\text{tot}} \cdot \frac{
    \left(\frac{Rp_{\text{max}}}{Rp_i}\right)^{Rp_{\text{exp}}}
}{N_P 
    \displaystyle\sum_{k=1}^{N_{Rp}} \left(\frac{Rp_{\text{max}}}{Rp_k}\right)^{Rp_{\text{exp}}} 
}
\end{equation}
Here $f_{i,j}$ is the re-binned stellar population for radius bin $i$ and period bin $j$, $F_{\text{tot}}$ is the total TRILEGAL population for the stellar class, $Rp_{\text{max}}$ is the maximum planet radius bin, $Rp_i$ is the planet radius of bin $i$, $Rp_{\text{exp}}$ is a scaling exponent parameter, $N_{Rp}$ is the number of radius bins, and $N_P$ is the number of period bins.  

For each planetary radius-period grid cell, the stellar populations $f_{i,j}$ are multiplied by the element-wise product (Hadamard) of the exoplanet occurrence rate from \cite{hsu2019} and \cite{dressing2015} with the detection probability for $k \geq 3$ transits (equation \ref{eqn:transprobk}).   

\subsection{Transit Confirmation Rules}
\label{sec:methodyield}
We now apply three rules for the confirmation of a transiting planet: 
\begin{description}
    \item[Star magnitude] The apparent magnitude of the host star must be in the range $16.5 \leq g \leq 25.5$  consistent with previous studies \citep{jacklin2015}.  The bright limit avoids saturation, and the faint limit corresponds to the sensitivity of the Simonyi Survey Telescope and the Rubin/LSST Camera for 30-second exposures  \citep{lsstsciencebook2009}. 
    \item[Transit Signal-to-Noise Ratio (SNR)] We used an SNR value of 5.0 where SNR is the ratio of transit depth to photometric uncertainty in Table~\ref{tab:rubinerr}.   
    \item[Number of transits] We require at least three observed transits to occur in at least one 30-second exposure during the 10-year survey duration.  This is a heuristic requirement to give confidence that periodic dips are present and thereby reduce False Alarms. A similar requirement was set by \citet{borucki2010} and \citep{christiansen2025} for Kepler detection. 
\end{description}

These three transit confirmation rules are then applied to reduce $N_{trans,prelim}$ to the final number of potentially observable transiting planets in each radius-period bin, $N_{trans}(i,j)$.  (The cautionary term "potentially" is used because other constraints will be discussed in \S\ref{sec:results}.) The sample $N_{trans,prelim}(i,j)$ is truncated by the magnitude range $16.5<g<25.5$ and the transit depth $SNR \geq 5$.  Targets with $k<3$ transits in the 10~year survey are removed.  The $N_{trans}(i,j)$ counts represent the main results reported in the next section.  

\section{Transit Observation Results}
\label{sec:results}
\subsection{Potentially Observable Transits from the WFD and DDF Surveys}
\label{sec:detectable}
The WFD survey will observe each field twice every third night over a ten-year period, with each exposure lasting 30 seconds and separated by approximately 30 minutes. According to Keplerian orbital mechanics, exoplanet transits typically last at least an hour, even for M-class stars. With only two exposures spaced 30 minutes apart, the survey cannot capture both the ingress and egress phases of a transit. Observing both phases is essential to reliably attribute a brightness dip to an exoplanet transit rather than to other sources of variability. Therefore, the cadence of the WFD survey is generally insufficient for the reliable confirmation of exoplanet transits.

\begin{figure}
\centering
\includegraphics[width=0.99\textwidth]{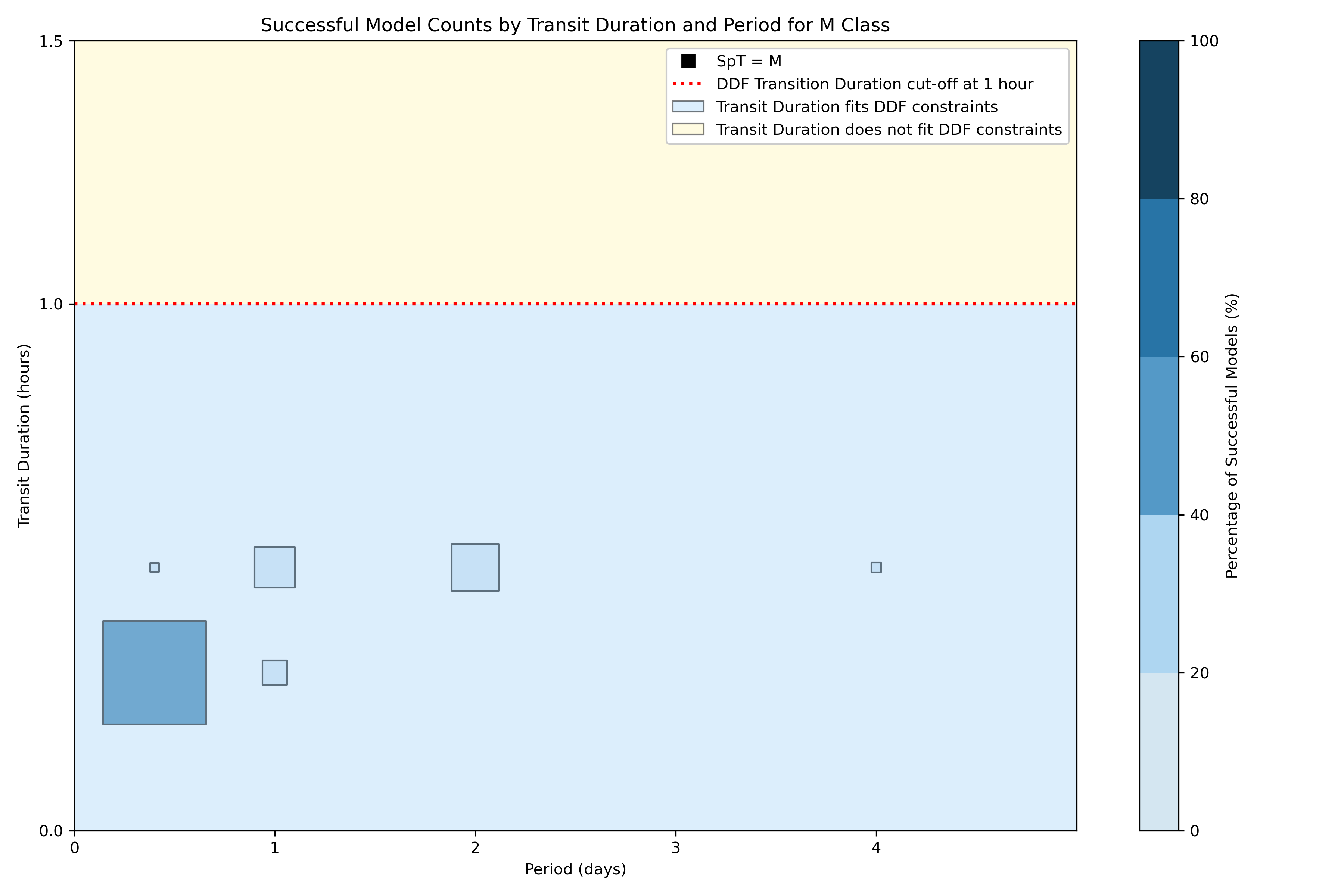}
\caption{This figure uses data from the DDF survey and highlights the results for M-class star models. F-, G-, and K-class star models fall only into the upper portion of the plot. WFD models are not represented because none of the WFD models pass the criteria for exoplanet detection. The dotted red line represents the maximum dwell time for the DDF survey. The size of the squares indicates that most of the successful confirmations will be for orbital periods of less than 3 days. A few models with orbital periods around 4 days are also possible.}
\label{fig:populations}
\end{figure}

However, the 10-year DDF survey  will be successful in confirming a small number of transits. Figure~\ref{fig:populations} overlays the parameters for  synthetic models on the grid developed by \cite{dressing2015}. This representation highlights the challenges of exoplanet confirmations in the DDF survey in that the successful models are only seen in one section of the grid. The values represented are from the models built for the six DDF fields with denser cadences than the WFD survey.

\begin{deluxetable*}{lrrrrrrrrrr}
\tablecaption{Probability of Exoplanet Confirmation in DDF by Period and Radius}
\label{tab:prob_bin}
\tablehead{& \multicolumn{10}{c}{Orbital Period (days)} \\
\colhead{$R_{\oplus}$} & \colhead{1.0} & \colhead{2.0} & \colhead{4.0} & \colhead{8.0} & \colhead{16.0} & \colhead{32.0} & \colhead{64.0} & \colhead{128.0} & \colhead{256.0} & \colhead{500.0} }
\startdata                                                                     
4.0  &  0.0086 & 0.0086 & 0.0086 & 0.0086 & 0.0086 & 0.0086 & 0.0086 & 0.0086  & -  & -  \\
3.5  &  0.0086 & 0.0086 & 0.0086 & 0.0086 & 0.0086 & 0.0086 & 0.0086 & 0.0086  & -  & -  \\
3.0  &  0.0086 & 0.0086 & 0.0086 & 0.0086 & 0.0086 & 0.0086 & 0.0086 & 0.0086  & -  & -  \\
2.5  &  0.0085 & 0.0085 & 0.0085 & 0.0085 & 0.0085 & 0.0085 & 0.0085 & 0.0085  & -  & -  \\
2.0  &  0.0085 & 0.0085 & 0.0085 & 0.0085 & 0.0085 & 0.0085 & 0.0085 & 0.0085  & -  & -  \\
1.5  &  0.0083 & 0.0083 & 0.0083 & 0.0083 & 0.0083 & 0.0083 & 0.0083 & 0.0083  & -  & -  \\
1.0  &  0.0076 & 0.0076 & 0.0076 & 0.0076 & 0.0076 & 0.0076 & 0.0076 & 0.0076  & -  & -  \\
\enddata
\tablecomments{These probabilities were derived as averages from models that passed the basic confirmation factors of apparent magnitudes within the constraints of the Rubin Observatory, transit depths that are at least five times the value of the Gaussian noise attributable to the telescope and at least three observing periods during the course of the ten year mission.}
\end{deluxetable*}

\begin{deluxetable*}{lrrrrrrrrrr}
\tablecaption{Weighted Distribution of Stellar Population for M-dwarfs in DDF}
\label{tab:pop_bin}
\tablehead{& \multicolumn{10}{c}{Orbital Period (days)} \\
\colhead{$R_{\oplus}$} & \colhead{1.0} & \colhead{2.0} & \colhead{4.0} & \colhead{8.0} & \colhead{16.0} & \colhead{32.0} & \colhead{64.0} & \colhead{128.0} & \colhead{256.0} & \colhead{500.0} }
\startdata                                                                     
4.0   &   3933  & 3933  & 3933  & 3933  & 3933  & 3933  & 3933  & 3933  & 3933  & 3933 \\
3.5   &   3942  & 3840  & 3741  & 3645  & 3551  & 3459  & 3370  & 3283  & 3198  & 3119 \\
3.0   &   3952  & 3736  & 3531  & 3338  & 3155  & 2983  & 2819  & 2665  & 2519  & 2386 \\
2.5   &   3964  & 3616  & 3298  & 3008  & 2744  & 2503  & 2283  & 2082  & 1899  & 1738 \\
2.0   &   3979  & 3474  & 3034  & 2649  & 2313  & 2019  & 1763  & 1540  & 1344  & 1179 \\
1.5   &   3998  & 3300  & 2724  & 2248  & 1855  & 1531  & 1264  & 1043  &  861  &  715 \\
1.0   &   4025  & 3069  & 2340  & 1784  & 1360  & 1037  &  791  &  603  &  460  &  354 \\
\enddata
\tablecomments{Using Equation \ref{eqn:popweighted}, we distributed the entire population (191352) of M class stars for the DDF survey.} 
\end{deluxetable*}

Only planets with transit durations shorter than the approximate 1-hour dwell time of the intra-night DDF cadence will have both ingress and egress observed. Capturing both the start and end of the transit is crucial for determining the full shape of the light curve dip, which in turn helps distinguish genuine exoplanet transits from other sources of variability. Planets with longer transits may produce detectable dips, but without clear ingress and egress measurements, these events are more likely to be confused with other types of stellar variability or systematic noise. In contrast, transits that are fully sampled within the cadence window—showing both ingress and egress—can be more reliably identified and confirmed as planetary in origin, reducing the risk of false positives.

Our results, as illustrated in Figure \ref{fig:populations}, indicate that, given current Rubin/LSST observing strategies, exoplanet transit confirmations over 10 years will be largely limited to short-period planets around M dwarfs in DDF fields. This limited yield is primarily due to the cadence and observational constraints of the Rubin/LSST mission. In the DDF survey, we modeled approximately 0.5 million M-dwarf transit scenarios and found that 73\% were rejected by these constraints. Of the rejected models, the vast majority (99\%) failed because their apparent magnitudes fell outside Rubin's detection limits. Additionally, 43\% did not reach the required SNR of 5, and 2\% lacked at least three transit observations over the 10-year survey.

\subsection{Further Constraints on Transit Observations in LSST Surveys}
\label{methodsver}
Two major issues can impede the reliable observation of planetary transits that have not been treated in the analysis so far. First, the light curve includes only ideal Gaussian noise, missing possible instrumental and stellar variations.  These could include instrumental red noise, image blending, seeing variations, and many types of stellar variability.  Variations due to magnetic activity are most commonly seen in Kepler and TESS light curves: evanescent and rotationally modulated star spots, granulation, and flares. The simulation of these effects is complex \citep[e.g.][]{Sulis2023} and is beyond the scope of this study.  The Kepler mission found that several percent of stars have inter-quartile ranges exceeding 0.1~mag and roughly half have ranges exceeding 0.001~mag \citep[Fig 15][]{Caceres2019b}.  Table~\ref{tab:rubinerr} shows that this is comparable to the standard deviation of Gaussian noise for brighter LSST stars with $g \gtrsim 17-18$.  As Kepler stars are mostly F-G-K type spectral types, the M stars of the DDF fields are likely to have even higher magnetic activity variability. The effects of magnetic activity may thus become significant at fainter magnitudes. 

A second limitation to confirming potential detectable transits in \S\ref{sec:detectable} concerns the interpretation of periodograms.  A periodogram plots the strength of a periodic signal for a large number of independent frequencies.  The Lomb-Scargle periodogram (LSP) generalization of the classical Fourier Schuster periodogram is not appropriate for planetary transit discovery because it assumes a continuous sinusoidal signal rather than a periodic sequence of brief dips in brightness.   The Box-Least Squares (BLS) algorithm \citep{Kovacs2002} has been most widely used for light curves with regular cadence; the Transit Comb Filter is similar but used for light curves where stellar variations are detrended with the differencing operator \citep{Caceres2019a}.  When the cadence interval is shorter than the ingress and egress time scale, as with the forthcoming PLATO satellite, the Transit Least Squares (TLS) algorithm that accounts for stellar limb darkening has better performance \citep{hippke2019, heller2019b, heller2019a, heller2022}.  A nonparametric algorithm such as Phase Dispersion Minimization \citep{Stellingwerf1978} can also be used.  

Irrespective of the algorithm, any periodogram can suffer a variety of difficulties including spectral leakage, aliasing, deterioration at periods close to the survey duration, and non-Gaussian distributions of powers even when applied to light curves with Gaussian noise.  Alias structures arising from light curves with diurnal gaps, which are unavoidable for Rubin Observatory surveys, can be complex \citep{VanderPlas2018}.  Non-Gaussian structures might be treated using statistical methods of Extreme Value Theory \citep{Baluev2008}.   It is thus possible that a `potentially' detectable period identified in \S\ref{sec:detectable} will be practically difficult to confidently identify in the chosen periodogram.   

To examine this possible difficulty, we initially examined the LSP and BLS period-search algorithms to identify periodic transit signals in light curves identified in \S\ref{sec:detectable}.  These techniques proved to be insufficient for robust transit observation in DDF light curves.  The periodograms have reduced sensitivity to shallow, short-duration signals and to the observing window limits imposed by the DDF observing strategy.  We then turned to the TLS algorithm that is optimized to find shallow, periodic transit-like signals in noisy and unevenly sampled light curves.

\subsection{TLS-Based Characterization of DDF Potential Transits}
\label{sec:resultsdet}
\begin{figure}   
\centering
\includegraphics[height=0.25\textheight]{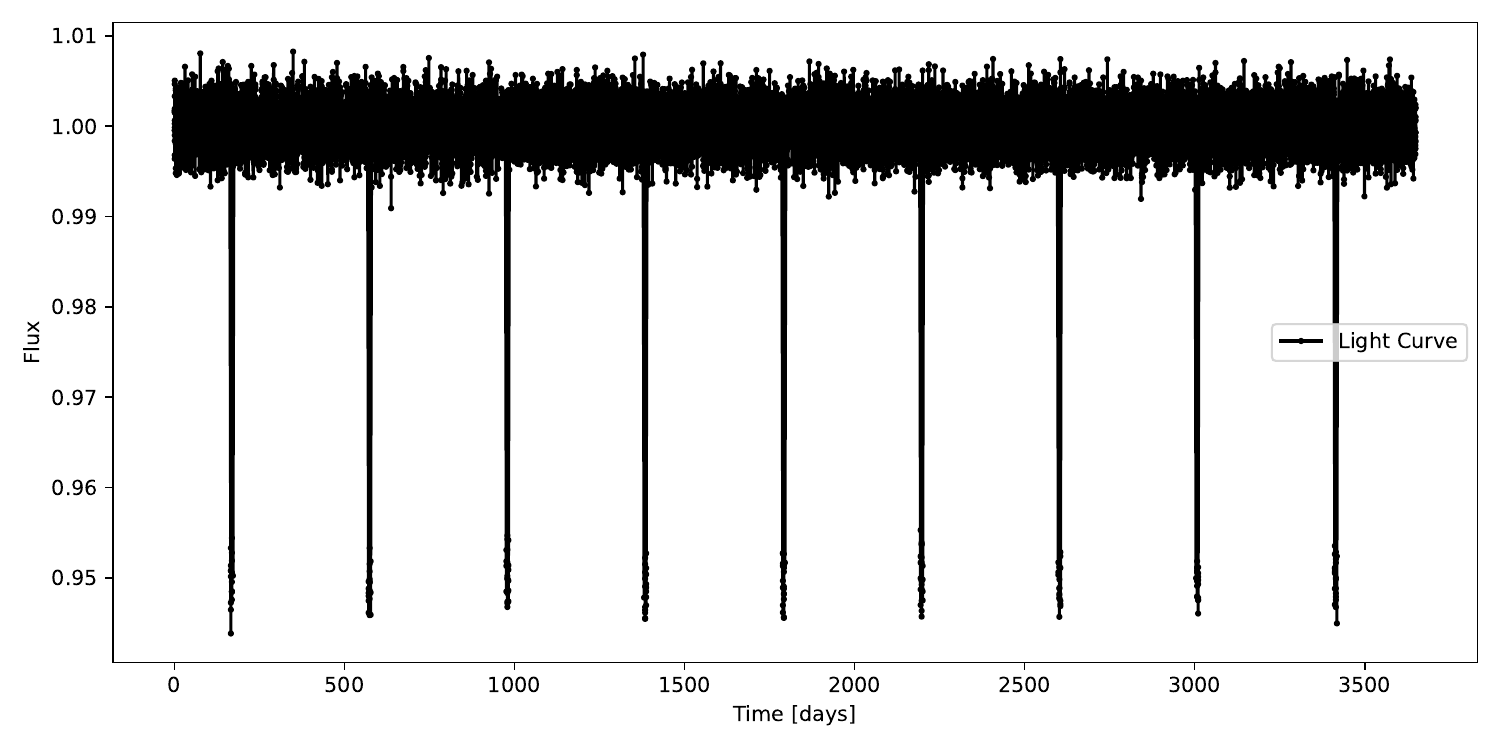} \\
\includegraphics[height=0.25\textheight]{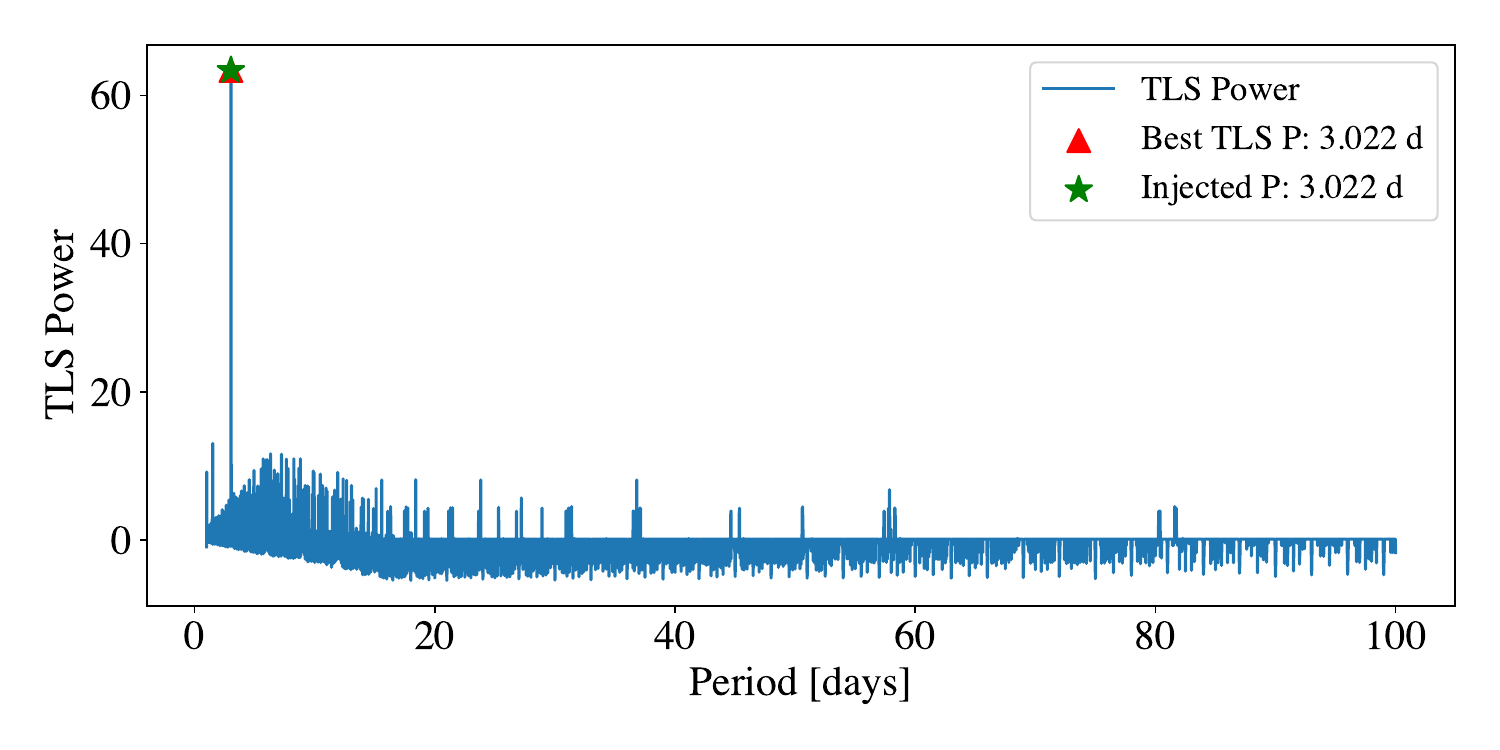}\\
\includegraphics[height=0.25\textheight]{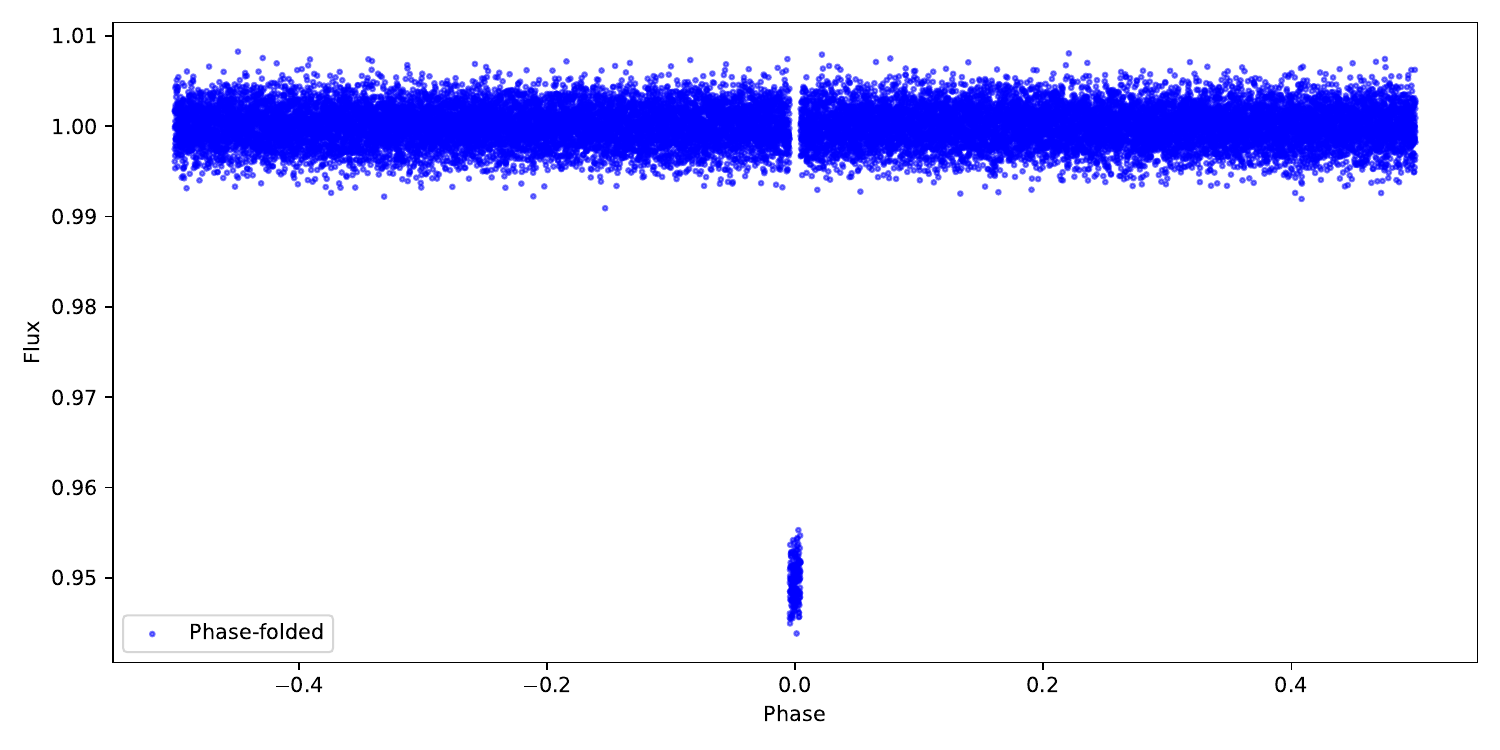}
\caption{Top: Ten year synthetic DDF light curve for an exoplanet orbiting an M dwarf. This modeled planet has a radius of 2.5 Earth radii and a period of 3.022 days. The observable transits are only those transits that can be observed during the Rubin/LSST mission under the defined observational and cadence strategy.  Middle: TLS periodogram of the full 3650 day DDF survey. The green star marks the injected transit mid-point while the red triangle indicates the best-fit orbital period identified by TLS. Bottom: Phase-folded light curve for the example system using the period identified by TLS.} 
\label{fig:periodograms}
\end{figure}

Figure~\ref{fig:periodograms} shows a simulated light curve in which the TLS periodogram precisely identifies the injected signal. Although the TLS algorithm performed well on these synthetic light curves, these results are optimistic, since our simulations omit instrumental and stellar variations. 

For the models that met the three elimination criteria discussed in \S\ref{sec:detectable}, TLS provides 100\% correct characterization of periods over the ten year mission. Orbital periods of 3 days or less can also be recovered in less than the full ten-year mission. For example, the period for the model illustrated in Figure \ref{fig:periodograms} was recovered by TLS with data from 200 observation days, which is equivalent to 600 calendar days or less than 2 years. 

Our analysis (Figure~\ref{fig:periodograms}) shows that periodic signals can be detected in simulated light curves that meet all criteria, even at lower SNRs. However, testing the TLS process using TESS data rather than our synthetic data suggests that, for Rubin/LSST to be effective for exoplanet transit searches, further advances—such as improved transit centering or machine learning classifiers—will be required to boost detection rates and reduce false positives. 

\subsection{Final Results for the DDF Survey}
Table \ref{tab:per_bin} summarizes our results for the 10-year DDF survey, incorporating all relevant factors discussed above: LSST survey sensitivity (\S\ref{sec:LSSTsurveys}), TRILEGAL stellar populations in the DDF Fields (\S\ref{sec:star_pop}), planet occurrence rates from the Kepler mission (\S\ref{sec:planet_pop}), transit detection probabilities (\S\ref{sec:planet_pop}), our transit detection criteria (\S\ref{sec:methodyield}), and TLS periodogram successful charterization rates (\S\ref{sec:resultsdet}). The predicted number of detected transiting planets is shown in column 3. We expect to recover $\approx{200}$  transiting planets, most of which are short-period ($P < 4$ days) planets with radii $R \simeq 2$~R$_\oplus$ orbiting early-M (M0V–M5V) main sequence stars. These predictions are optimistic because they do not account for instrumental or intrinsic stellar variability.

\begin{deluxetable*}{lrr}
\tablecaption{Estimate of Exoplanets Detectable within the 10-year DDF survey}
\label{tab:per_bin}
\tablehead{\colhead{Class} & \colhead{\# Obs} & \colhead{\# Rec}  }
\startdata
M0V &  19780  & 23.72  \\
M1V &  23111  & 27.72   \\
M2V &  24666  & 29.58  \\
M3V &  42363  & 50.81   \\
M4V &  40131  & 48.13   \\
M5V &  28028  & 33.61   \\
M6V &  10191  & 12.22   \\
M7V &  1838  & 2.20   \\
M8V &  898  & 1.08   \\
M9V &  288  & 0.35   \\
M9.5V &  58  & 0.07  \\
Total &  191352  & 229.49   \\
\enddata
\tablecomments{\# Obs: Estimated number of stars observed. \# Rec: Expected exoplanets recovered around M dwarfs over 10 years.}

\end{deluxetable*}

\section{Discussion} \label{Discussion}
\subsection{Summary}
We synthesize discrete models of exoplanetary systems and simulate their detectability under each Rubin/LSST survey scenario to forecast the surveys' potential for transit exoplanet science. Our approach accounts for stellar and planetary populations in survey areas, telescope sensitivity, survey observing cadence, with the requirement that a full transit (ingress to egress) be observed at least 3 times with $SNR \geq 5$ during the 10-year Rubin/LSST mission.  This framework enables a systematic assessment of expected exoplanet yields from wide-field surveys and highlights the principal limitations imposed by cadence and operational constraints.

Except for a handful of hot planets orbiting faint M0-M5 main sequence stars detectable in DDF fields, no transits will be confidently confirmed.  The principal difficulty is the sparse, multiband observing strategy that prioritizes cosmology and extragalactic science. These cadence policies prevent significant galactic planetary transit confirmations. 

\subsection{Comparison with Previous Work}
\citet{lund2015}, \citet{jacklin2015} and \citet{jacklin2017} report successful detection of hot Jupiters with periods $0.5-20$~days around faint G stars from the WFD survey.  The DDF survey would be more successful, recovering $>30$\% of sub-Saturn-size exoplanets with periods up to 20~days.  Some detections would be made within the first $1-2$~years of the surveys, while characterization fractions improve with the duration of the survey. 

Building on these results, our analysis incorporates more recent insights into the challenges of exoplanet detection and confirmation with Rubin/LSST cadence, particularly focusing on the effects of sparse sampling, window functions, and period aliasing. Previous studies, such as \citet{jacklin2015, jacklin2017}, carefully defined their detection and recovery criteria and discussed the impact of window functions, phase coverage, and false positive thresholds. Our simulations extend this work by explicitly modeling the confirmation process under the latest survey strategies and highlight the specific difficulties of period recovery for planets with longer periods or smaller radii given the expected cadence and noise environment of Rubin/LSST.

Although algorithms such as the Box-Least Squares (BLS) method have been effective in identifying transit signals from ground-based surveys even with sparse or noisy data \citep{cameron2006}, our results suggest that, under the planned Rubin/LSST cadence, the recovery of transiting planets is likely to be limited except in favorable cases (e.g., larger planets orbiting early-M dwarfs). This is primarily due to the combination of cadence, window function effects, and the relatively low signal-to-noise ratio expected for small planets.

We therefore emphasize that, while the methodologies and detection pipelines developed in prior work remain powerful tools, the specific observational constraints of Rubin/LSST may limit the yield of confirmed transiting exoplanets relative to earlier projections, unless additional follow-up or probabilistic validation techniques are employed.

\subsection{Opportunities for Enhanced Yield}
Our analysis identifies several strategies that would improve exoplanet detection prospects with the Rubin Observatory:
\begin{description}
    \item[Targeting the Galactic disk] Establishing a DDF field on the Galactic disk, while maintaining the DDF cadence, could increase the detection yield by a factor of $\gtrsim 50$ compared to current DDF fields.
    \item[Increased observing cadence] Implementing a daily DDF cadence in such a disk field could increase detection rates by $\sim 5000$ over current DDF detection rates.
    \item[Algorithmic improvements] Enhanced detection algorithms, including machine learning or improved transit centering, could further increase detection rates, reducing false alarms and false positives.
    \item[Multi-band observations] Once a candidate exoplanet is found, returning with multiple band passes could test achromaticity and  improve sensitivity for stars whose peak luminosity is outside the $g$-band.
    \item[White-dwarf hosts] The DDF cadence may be aligned with the brief transit durations anticipated for planets orbiting white dwarfs. We recommend that this be studied in later work.  
    \item[Gaia satellite] For candidates with magnitudes $g \lesssim 21$, cross-validation with Gaia variability and astrometric data would be valuable. 
\end{description}

The most compelling path forward is to implement a dedicated micro-survey targeting a single densely populated Galactic field, utilizing intensive cadences over a continuous time block. As proposed by \citet{Feigelson2023}, even a single-band, single-field, single-night micro-survey has the potential to yield valuable confirmations. However, extending this strategy to encompass multiple seasons of sequential nights—repeated at regular intervals throughout the Rubin/LSST mission—would vastly increase sensitivity to a much broader diversity of host stars and exoplanet properties. Such a campaign—uniquely enabled by the Rubin Observatory’s combination of wide field, high cadence, and unprecedented photometric precision—would open a new window into exoplanetary demographics, capturing both common and rare planetary systems that remain inaccessible to other surveys. By pushing the boundaries of current methodologies, this approach promises to deliver transformative insights into planet formation and evolution, and to inform future missions targeting habitable worlds.

\section{Acknowledgements}
NSF–DOE Vera C. Rubin Observatory, funded by the U.S. National Science Foundation and the U.S. Department of Energy's Office of Science, will perform the Legacy Survey of Space and Time using the Rubin/LSST Camera and the Simonyi Survey Telescope. Rubin Observatory is a joint program of NSF NOIRLab and DOE’s SLAC National Accelerator Laboratory.

This research uses services or data provided by the Astro Data Lab, which is part of the Community Science and Data Center (CSDC) Program of NSF NOIRLab. NOIRLab is operated by the Association of Universities for Research in Astronomy (AURA), Inc. under a cooperative agreement with the U.S. National Science Foundation.

This research has used the NASA Exoplanet Archive, which is operated by the California Institute of Technology, under contract with the National Aeronautics and Space Administration under the Exoplanet Exploration Program.

The authors acknowledge Research Computing at Arizona State University for providing HPC and storage resources that have contributed to the research results reported in this article.Computations were performed on the Sol supercomputer at Arizona State University \citep{HPC:ASU23}.  
\bibliography{references.bib}
\end{document}